\documentclass[conference]{IEEEtran}
\IEEEoverridecommandlockouts
\usepackage{amsmath,amssymb,amsfonts}
\usepackage{algorithmic}
\usepackage{graphicx}
\usepackage{textcomp}
\usepackage[dvipsnames]{xcolor}
\usepackage{hyperref}
\usepackage{balance}
\usepackage{booktabs}
\usepackage{caption}
\usepackage{inconsolata}
\usepackage{microtype}
\usepackage[numbers, sort&compress]{natbib}

\usepackage[scale=.9]{roboto}
\usepackage[scale=.9]{roboto-mono}
\usepackage{FiraSans}
\usepackage[T1]{fontenc}

\usepackage[noabbrev,capitalise]{cleveref}
\usepackage{syntax}

\usepackage{tikz}
\usetikzlibrary{positioning,shapes,trees,arrows}
\usetikzlibrary{arrows.meta}
\usepackage{tikz-qtree}
\usepackage[framemethod=tikz]{mdframed}
\usepackage{listings}
\usepackage[frozencache]{minted}
\setminted{
    numbers=left,
    xleftmargin=10pt,
    fontsize=\footnotesize,
}

\def\checkmark{\tikz\fill[scale=0.4,fill=OliveGreen](0,.35) -- (.25,0) -- (1,.7) -- (.25,.15) -- cycle;} 
\def\cross{\tikz [x=1.4ex,y=1.4ex,line width=.4ex, red] \draw (0,0) -- (1,1) (0,1) -- (1,0);}

\BeforeBeginEnvironment{minted}{\begin{mdframed}[backgroundcolor=gray!5,linecolor=black!50,linewidth=1pt,roundcorner=5pt]}
\AfterEndEnvironment{minted}{\end{mdframed}}

\lstnewenvironment{RISElisting}[1][]
  {%
   \mdframed[backgroundcolor=gray!5,linecolor=black!50,linewidth=1pt,roundcorner=5pt]%
   \lstset{#1}%
  }
  {\endmdframed}
\newcommand\Comment{\hfill\itshape\color{black!60}}

\newcommand{\RISE}{\robotoLight{R}\robotoRegular{I}\robotoMedium{S}\robotoBold{E}}
\newcommand{\Rise}{\robotoRegular{RISE}}
\newcommand{\DPIA}{\textsc{\robotomonoRegular{DPIA}}}
\newcommand{\Shine}{\robotomonoRegular{Shine}}
\newcommand{\Elevate}{\textsc{\firalight Elevate}}
\newcommand{\Lift}{\textsc{Lift}}

\newcommand{\Accep}[1]{\mathcal{A}}
\newcommand{\Cont}[1]{\mathcal{C}}

\begin{document}

\title{
\RISE{}\,\&\,\Shine{}:
Language-Oriented Compiler Design
}

\author{
    \IEEEauthorblockN{
        Michel Steuwer\IEEEauthorrefmark{1}
        Thomas K\oe{}hler\IEEEauthorrefmark{2}
        Bastian K\"opcke\IEEEauthorrefmark{3}
        Federico Pizzuti\IEEEauthorrefmark{1}
    }
    \IEEEauthorblockA{
        \IEEEauthorrefmark{1}\textit{University of Edinburgh, Scotland, UK}\quad
        \IEEEauthorrefmark{2}\textit{University of Glasgow, Scotland, UK}\quad
        \IEEEauthorrefmark{3}\textit{University of M\"unster, Germany}
            \\
        Email:
            \href{mailto:michel.steuwer@ed.ac.uk}{michel.steuwer@ed.ac.uk},
            \href{mailto:thomas.koehler@thok.eu}{thomas.koehler@thok.eu},
            \href{mailto:bastian.koepcke@wwu.de}{bastian.koepcke@wwu.de},
            \href{mailto:federico.pizzuti@ed.ac.uk}{federico.pizzuti@ed.ac.uk}
    }
}


\maketitle
\thispagestyle{plain}
\pagestyle{plain}

\begin{abstract}
The trend towards specialization of software and hardware --- fueled by the end of Moore's law and the still accelerating interest in domain-specific computing, such as machine learning --- forces us to radically rethink our compiler designs.
The era of a universal compiler framework built around a single one-size-fits-all intermediate representation (IR) is over.
This realization has sparked the creation of the MLIR compiler framework that empowers compiler engineers to design and integrate IRs capturing specific abstractions.
MLIR provides a generic framework for SSA-based IRs, but it doesn't help us to decide how we should design IRs that are easy to develop, to work with and to combine into working compilers.

To address the challenge of IR design, we advocate for a \emph{language-oriented compiler design} that understands IRs as formal programming languages and enforces their correct use via an accompanying type system.
We argue that programming language techniques directly guide extensible IR designs and provide a formal framework to reason about transforming between multiple IRs.
In this paper, we discuss the design of the \Shine{} compiler that compiles the high-level functional pattern-based data-parallel language \Rise{} via a hybrid functional-imperative intermediate language to C, OpenCL, and OpenMP.

We compare our work directly with the closely related pattern-based \Lift{} IR and compiler.
We demonstrate that our language-oriented compiler design results in a more robust and predictable compiler that is extensible at various abstraction levels.
Our experimental evaluation shows that this compiler design is able to generate high-performance GPU code.
\end{abstract}


\section{Introduction}
\label{sec:intro}
The end of Moore's law and Dennard's scaling has opened the door to \emph{a new golden age for computer architecture}~\cite{DBLP:journals/cacm/HennessyP19} that is characterized by specialization of hardware, which we already see in the diverse landscape of AI and ML accelerators.

Together with the specialization in hardware, we observe an accelerating trend of specialized software solutions as well, again most prominently in the domain of machine learning with popular frameworks such as TensorFlow and PyTorch.

To exploit the tremendous efficiency gains of specialized hardware, we need \emph{specialized compilers} capable of translating the specialized software to the hardware by employing optimizations depending on the software domains and hardware capabilities.
Gone are the days of a one-size-fits-all compiler with a single intermediate representation (IR) sufficient to optimize for the few standard hardware architectures of interest.

These observations sparked the creation of the MLIR~\cite{DBLP:conf/cgo/LattnerAB0DPRSV21} compiler framework as a natural evolution of the well-known LLVM~\cite{DBLP:conf/cgo/LattnerA04} compiler framework that embraced the idea of a single static assignment (SSA)-based IR and with it managed to unify community development efforts in the compiler community at an unprecedented scale.

MLIR --- in contrast to LLVM --- embraces diversity and specialization of IRs and their optimizations.
With only few core constraints on the IR designs, most notably the insistence on SSA as a formal foundation, MLIR has already proven to be a versatile framework for expressing various IRs\footnote{\url{https://mlir.llvm.org/docs/Dialects/}}.
This includes traditional generic IRs (such as the LLVM IR itself),
IRs reflecting specific programming models (such as \texttt{OpenMP} and TensorFlow),
IRs specialized towards particular styles of optimizations (such as the polyhedral model),
IRs reflecting specific hardware features (such as vectorization and GPU-specific features),
all the way to IRs for hardware circuit design.

But besides its core foundations, MLIR provides little guidance on how to design \emph{good} IRs that are easily extended and easy to work with, which are robust by ruling out ill-formed programs, and which compose well when integrated in full compilers.
MLIR provides a framework for designing compilers as compositions of various SSA-based IRs, but what should be our mental framework for designing the IRs themselves?

In this paper, we argue that we should look at programming language design for inspiration.
Of course, this observation is not new, but we believe that this paper adds a new perspective by providing insights into our experience of following this approach.
We present our experience of designing the optimizing \Shine{} compiler for \Rise{} --- a functional data-parallel programming language in the spirit of \Lift{}~\cite{DBLP:conf/icfp/SteuwerFLD15,DBLP:conf/cgo/SteuwerRD17}.
\Rise{} builds on a small functional core language that is enriched with data-parallel patterns such as \texttt{map} and \texttt{reduce} suitable for expressing numerical computations over tensors relevant in many domains ranging from machine learning to scientific computing.
The functional and pattern-based design makes it easy to exploit high-level optimization opportunities via rewriting and to generate high-performance code as explored before in the context of \Lift{}~\cite{DBLP:conf/ppopp/RemmelgLSD16,DBLP:conf/cgo/HagedornSSGD18} and \Rise{}~\cite{DBLP:journals/pacmpl/HagedornLKQGS20,DBLP:conf/cgo/KoehlerS21}.

To summarize, this paper makes the following contributions:
\begin{itemize}
    \item We advocate a language-oriented IR and compiler design by reporting our experience of engineering the \Shine{} compiler and discussing its design (\cref{sec:overview});
    \item we describe the benefits of our compiler design: a clear separation of concerns between optimizing and code generation (\cref{sec:optimizing-vs-lowering}), formalizing of invariants and assumptions about IRs in type systems (\cref{sec:type-system}), and its extensibility at various levels in the compiler (\cref{sec:extensibility});
    \item we present an experimental evaluation demonstrating the quality of the generated code (\cref{sec:evaluation}).
\end{itemize}

\section{Background in IR and Compiler Design}
\label{sec:background}
Compiler IRs have been studied extensively before.
In the community of optimizing compilers for imperative languages, SSA~\cite{SSA-Book} --- developed in the late 1980s at IBM research~\cite{DBLP:conf/popl/RosenWZ88,DBLP:conf/popl/AlpernWZ88} --- has by now been established as the standard form for IRs.
SSA makes \emph{def-use-chains} explicit and guarantees that \emph{def}s are introduced before they are \emph{use}d.
Since the mid-1970s, the functional community used lambda calculus (introduced in the 1930s by Alonzo Church) as a compiler IR.
The \emph{Lambda Papers} series even advocated the flexibility of lambda calculus to model non-functional languages~\cite{lambda-imperative, lambda-declarative}.

In the mid/late 1990s \citeauthor{DBLP:conf/irep/Kelsey95} and \citeauthor{DBLP:journals/sigplan/Appel88} pointed out a direct correspondence between SSA and the functional lambda calculus representations~\cite{DBLP:conf/irep/Kelsey95,DBLP:journals/sigplan/Appel88}.
Despite this fundamental correspondence, the functional and imperative compiler communities have continued to develop largely independently.

The rise of machine learning in the last decade has sparked interest in domain-specific compilers and their IRs, specifically for numerical computations over tensors --- multidimensional arrays.
These specialized compilers such as TensorFlow~\cite{DBLP:conf/osdi/AbadiBCCDDDGIIK16}, PyTorch~\cite{DBLP:journals/corr/abs-1912-01703}, or TVM~\cite{DBLP:conf/osdi/ChenMJZYSCWHCGK18} often use a graph-based representation.
These systems have developed into complex optimizing compilers having taken on inspirations from related compilers such as Halide~\cite{DBLP:conf/pldi/Ragan-KelleyBAPDA13} and \Lift{}~\cite{DBLP:conf/icfp/SteuwerFLD15,DBLP:conf/cgo/SteuwerRD17}.
\Lift{} builds on a tradition of similar functional languages and compilers that existed before, such as SaC~\cite{DBLP:conf/ifl/Scholz96}.
More have emerged recently, including Accelerate~\cite{DBLP:conf/popl/ChakravartyKLMG11} and Futhark~\cite{DBLP:conf/pldi/HenriksenSEHO17} as well as Dex~\cite{DBLP:journals/pacmpl/PaszkeJDVRJRM21} by Google research.

An advantage of the functional representation is that optimizations are naturally expressed via rewriting allowing the automatic exploration of optimizations for example by stochastic search as explored by \Lift{}~\cite{DBLP:conf/icfp/SteuwerFLD15,DBLP:conf/ppopp/RemmelgLSD16}.
Stratego~\cite{DBLP:journals/scp/BravenboerKVV08} and more recently \Elevate{}~\cite{DBLP:journals/pacmpl/HagedornLKQGS20} show how to describe complex program transformations as composition of rewrites expressed in \emph{strategy languages} that provide a more formal treatment of the \emph{scheduling languages} found in Halide and TVM.

To accommodate the rising need for specialized IRs, MLIR~\cite{DBLP:conf/cgo/LattnerAB0DPRSV21} provides a common framework for designing IRs and transformations between them.
The different IRs, called \emph{dialects}, must follow the SSA form.
But in contrast to LLVM's IR, the awkward $\phi$-nodes are replaced with so-called \emph{regions} that allow for directly representing nested structures, such as control flow with multiple branches, e.g., an \texttt{if-then-else} operation.
MLIR allows for a broad range of IRs to be represented and provides itself little guidance on how to design dialects that are well-behaved, that is, dialects that:
1) providing a clear purpose;
2) have formally checked invariants and assumptions;
3) are easily extensible.

In this paper, we provide a perspective on how to design IRs that have these positive properties.
We share our experience in developing the \Shine{} compiler that composes a \Lift{}-like functional IR with a hybrid functional-imperative IR for compiling high-level programs to high-performance code.

\section[Shine: A Language-Oriented Compiler]{\Shine{}: A Language-Oriented Compiler}
\label{sec:overview}

\begin{figure}
    \centering
    \includegraphics[width=.9\columnwidth]{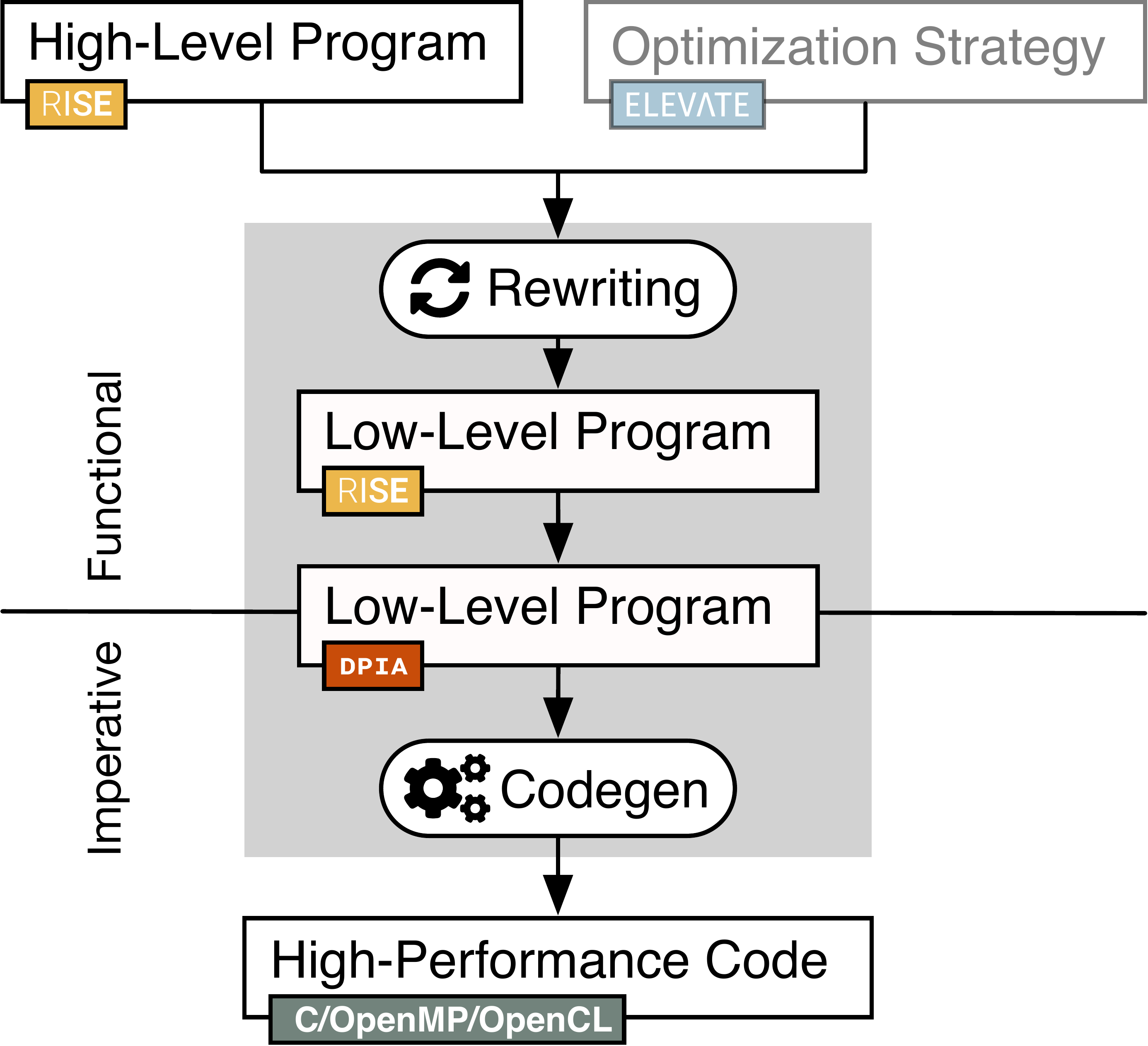}
    \caption{
        Overview of the language-oriented designed compiler \Shine{}:
        \emph{High-level programs} expressed in the functional language \Rise{} are rewritten to encode all major optimization and implementation choices.
        The resulting \emph{low-level program} is translated into \DPIA{} --- a language combining functional and imperative aspects, from which \emph{high-performance code} is generated.
        The language-oriented design results in a clear separation of concerns (optimizing vs. code generation), it simplifies the formalization of invariants and assumptions about the IRs in their type systems, and it's easily extensible by introducing abstractions at multiple levels in the compiler.
    }
    \label{fig:rise+shine-overview}
\end{figure}

\Cref{fig:rise+shine-overview} shows an overview of the \Shine{} compiler.
The overall design follows the compiler described by \citeauthor{DBLP:journals/pacmpl/HagedornLKQGS20} in~\cite{DBLP:journals/pacmpl/HagedornLKQGS20} where an optimization strategy written in a \emph{strategy language} (aka, scheduling language) called \Elevate{} specifies the optimizations to be applied to the input program.
\Elevate{} is described in detail in~\cite{DBLP:journals/pacmpl/HagedornLKQGS20} and we will not discuss it further in this paper, but we highlight that it allows for experts (human or machine) to directly control the optimizations to be performed.

The input program of the \Shine{} compiler is written in the functional data-parallel language \Rise{} that uses a restricted form of \emph{dependent types} to track the size of multi-dimensional arrays in the type.
This follows our language-oriented design principle in that this information is embedded into the type system of the IR rather than being represented as separate meta-data.
This has the main advantage that we have a formal account of the structure of the meta-data (here tracking array lengths) allowing us to ensure its consistency by the soundness of the type system, for example, the type system accounts for the basic fact that variables referred to in the types must be well-scoped.

The process of compiling a high-level \Rise{} program to high-performance code is shown in \cref{fig:rise+shine-overview}.
The process is broken into two steps, each with a corresponding language and type system serving as the IR.

First, the functional \emph{high-level program} is transformed via rewriting into a \emph{low-level program} that encodes optimization and implementation decisions.
The \Elevate{} optimization strategy, given as an input to the compiler, describes these decisions, such as:
Should we fuse multiple patterns to avoid intermediate buffers?
In which address space should we allocate intermediate buffers?
Should we perform \texttt{map} patterns in parallel or sequential?
In this rewriting step, choices for these decisions are directly encoded, resulting in a functional \emph{low-level program}.

The remaining steps translate the functional low-level program into an imperative low-level program.
For this, we introduce a new typed language called \DPIA{} as IR that combines functional and imperative aspects while preventing simultaneous sharing and mutation similar to Rust.
After translating the functional low-level \Rise{} program into a functional \DPIA{} program, we gradually translate it into an imperative program using a formal translation.
This translation preserves the implementation and optimization choices made before and does not make significant implementation choices itself.
Finally, \emph{high-performance code} is generated (\texttt{C}, \texttt{OpenMP}, or, \texttt{OpenCL} depending on the implementation choices made).

\section{Clearly Separating Optimizing from Code~Generation}
\label{sec:optimizing-vs-lowering}
One main principle observed in the \Shine{} compiler is the clear separation of IRs for optimizing and code generation.
In this design, two languages (\Rise{} and \DPIA{}) are composed, each serving as the IR for one task:
all optimizations and major implementation decisions are performed on \Rise{};
the code generation process of translating the functional to an imperative program is performed in \DPIA{}.

Keeping these two fundamental tasks separated has the key advantage that the IR for optimizing is kept simple and free of unnecessary low-level details.
These are only added after all optimization decisions are encoded in the functional program.

\begin{listing}[b]
\begin{RISElisting}[numbers=left,xleftmargin=10pt]
def mv = depFun((n: Nat, m: Nat) =>
  fun(M: Array[n, Array[m, f32]] =>
    fun(x: Array[m, f32] =>
      M |> map(fun(row =>
          zip(row)(x) |>
            map(fun(ax => fst(ax) * snd(ax))) |>
              reduce(add)(0.0f) )) )) )
\end{RISElisting}
\caption{
    Matrix-Vector-Multiplication expressed as a high-level program in \Rise{}.
    No optimization or implementation choices have been made yet, there are many ways to compile this program to high-performance code.
}
\label{fig:mm-rise}
\label{fig:rise-mv}
\end{listing}

To understand the optimization and code generation process of the \Shine{} compiler, we will follow an easy-to-understand running example: matrix-vector-multiplication (\cref{fig:mm-rise}).
This functional program describes the computation algorithmically without committing to a particular implementation strategy:
for each row of the matrix \texttt{M}, we compute the dot-product with the input vector \texttt{x}.
The dot-product is expressed in lines 5--7 by composition of the \texttt{zip} primitive that aligns the matrix row and the vector element wise, followed by the \texttt{map} primitive that multiplies each aligned pair, before the resulting array is summed up using the \texttt{reduce} primitive.
In \Rise{}, sizes of multi-dimensional arrays can be tracked symbolically in their types.
For this, we introduce a \emph{dependent function} in line 1 that abstracts over the type-level size values in the same way as ordinary functions abstract over computational values.
The introduced variables \texttt{n} and \texttt{m} are now allowed to appear in the \emph{types} of the nested expression.
In a non-language-oriented IR design, such information is often modeled as meta-data without a clear understanding of scoping rules and how it interacts with the computational operations of the IR.
Next, let us investigate the \Rise{} language itself.

\subsection[Rise expressions, types, and primitives]{\Rise{} expressions, types, and primitives}
\begin{figure}
    \centering
\begin{lstlisting}[basicstyle=\ttfamily\footnotesize, commentstyle=\itshape\color{black!60}, mathescape,language=c,escapeinside={!:}{:!},]
$E$ := x | 0.0f |               !:\Comment variables and literals:!
     fun(x => $E$) |            !:\Comment function abstraction:!
     $E$ |> $E$ | $E$($E$) |    !:\Comment function application:!
     depFun(x: $K$ => $E$) |    !:\Comment dependent fun. abstraction:!
     $E$($N$) | $E$($DT$) | $E$($A$) |     !:\Comment dependent fun. application:!
     map | reduce | reduceSeq | zip | ... !:\Comment primitives:!
\end{lstlisting}
\vspace{.5em}
\begin{tikzpicture}[grow'=right]
\tikzset{level distance=60pt,
         every tree node/.style={anchor=west},
         every node/.style={
            draw,
            shape=rectangle,
            rectangle split,
            rectangle split parts=2,
            rectangle split horizontal,
            rectangle split ignore empty parts,
            font=\footnotesize\ttfamily\bfseries,
            every two node part/.style={font=\ttfamily\footnotesize}
         },
         edge from parent/.style={
             draw, thick, {Stealth[length=5mm, open, round]}-,
             edge from parent path={(\tikzparentnode.east)
                                        -- +(+17pt,0)
                                        |- (\tikzchildnode.west)}
            },
        execute at begin node=\strut,
        }
\Tree [.\node[rectangle split horizontal=false] {
                \emph{Expr}
                \nodepart{two}
                t: Type
            };
    [.\node {
        Identifier
        \nodepart{two}
        name: String
    }; ]
    [.\node {
        Literal
        \nodepart{two}
        value: String
     }; ]
    [.\node {
        Lambda
        \nodepart{two}
        x: Identifier; e: Expr
     }; ]
    [.\node {
        Apply
        \nodepart{two}
        f: Expr; e: Expr
     }; ]
    [.\node {
        DepLambda<K,I>
        \nodepart{two}
        kind: K; x: I; e: Expr
     }; ]
    [.\node {
        DepApply<K,T>
        \nodepart{two}
        kind: K; f: Expr; x: T
     }; ]
    [.\node {\emph{Primitive}}; ]
]
\end{tikzpicture}
\caption{Grammar of \Rise{} expressions (above) and corresponding diagram of classes (below) representing them.}
\label{fig:rise-expr-overview}
\end{figure}
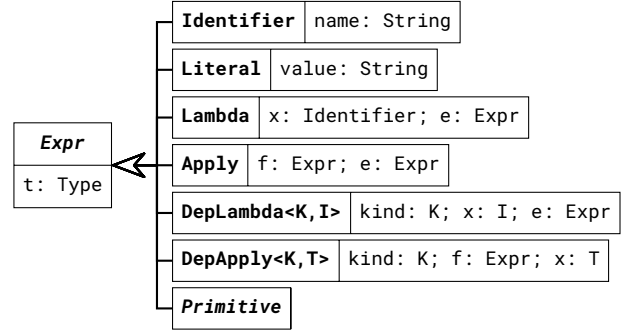

From a users' perspective, \Rise{} shares many common features with the \Lift{} IR, but their implementations differ significantly.
The \Lift{} implementation separates expressions and function declarations and lacks a formalized type system, in particular \Lift{} lacks the concept of a function type.
In contrast, \Rise{} unifies all language constructs under a single \texttt{Expr} superclass.
This simplified design follows the usual implementation of languages in the functional community by using an \emph{algebraic data type (ADT)} to represent the various language constructs.
ADTs model data in a structured way by organizing it in a finite set of alternative cases, where each case is represented as a \texttt{struct} containing multiple fields.
ADTs are a perfect fit to represent grammars, such as \Rise{}'s grammar shown at the top of \cref{fig:rise-expr-overview}.
Support for ADTs is directly built into many functional languages, but support has recently also been added in non-functional languages, e.g., they are known as \texttt{enum} in Rust and Swift and as \texttt{std::variant} in C++.
A straightforward way to encode ADTs in an object-oriented language is to build a flat class hierarchy, as shown at the bottom of \cref{fig:rise-expr-overview}.

The \texttt{Expr} class is abstract with seven subclasses modeling the language constructs, such as:
\texttt{Lambda} and \texttt{Apply} to model function abstraction (aka, lambdas) and application,
\texttt{DepLambda/Apply} to model function abstraction/application over variables at the type level (e.g, for reasoning about the length of arrays),
and finally, \texttt{Primitive} to represent primitives, such as \texttt{map} and \texttt{reduce}.

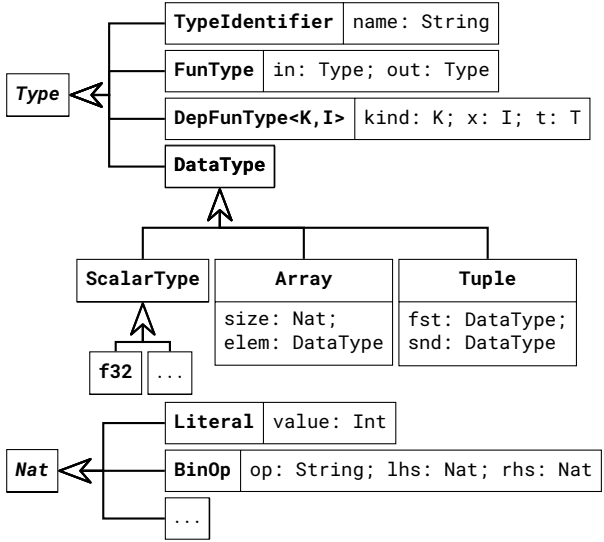
\begin{figure}
    \centering
\begin{lstlisting}[basicstyle=\ttfamily\footnotesize, commentstyle=\itshape\color{black!60}, mathescape,language=c,escapeinside={!:}{:!},]
$T$ := t | $DT$ |                  !:\Comment type variables \& data types:!
     $T$ $\rightarrow$ $T$ | (x: $K$) $\rightarrow$ $T$ !:\Comment function types:!
     
$DT$ := f32 | ... |                     !:\Comment scalar types:!
     Array[$N$,$DT$] | Tuple[$DT$,$DT$] !:\Comment array \& tuple types:!
     
$N$ := 0|1| ... | $N$ + $N$ | $N$ * $N$ | ... !:\Comment natural numbers:!
$A$ := Global | Local | Private !:\Comment OpenCL address spaces:!
$K$ := Nat | DataType | AddrSp !:\Comment kinds:!
\end{lstlisting}
\vspace{.5em}
\begin{tikzpicture}
\tikzset{every tree node/.style={anchor=west},
         every node/.style={
            draw,
            shape=rectangle,
            rectangle split,
            rectangle split parts=2,
            rectangle split horizontal,
            rectangle split ignore empty parts,
            font=\footnotesize\ttfamily\bfseries,
            every two node part/.style={font=\ttfamily\footnotesize}
         },
         edge from parent/.style={
             draw, thick, {Stealth[length=5mm, open, round]}-
            },
        execute at begin node=\strut,
        }
\begin{scope}[
    grow'=right,
    xshift=-2cm,
    edge from parent path={(\tikzparentnode.east) -- +(+15pt,0) |- (\tikzchildnode.west)},
    level distance=60pt
]
\Tree [.\node[rectangle split horizontal=false] {
                \emph{Type}
            };
    [.\node {
        TypeIdentifier
        \nodepart{two}
        name: String
     }; ]
    [.\node {
        FunType
        \nodepart{two}
        in: Type;
        out: Type
    }; ]
    [.\node {
        DepFunType<K,I>
        \nodepart{two}
        kind: K;
        x: I;
        t: T
     }; ]
    [.DataType ]
]
\end{scope}
\begin{scope}[yshift=-0.95cm, xshift=60pt,
    grow=down,
    edge from parent path={(\tikzparentnode.south) -- +(0, -15pt) -| (\tikzchildnode.north)},
    level distance=35pt,
    every tree node/.style={anchor=north},
    every node/.style={
            draw,
            shape=rectangle,
            rectangle split,
            rectangle split parts=2,
            rectangle split horizontal=false,
            align=left,
            rectangle split ignore empty parts,
            font=\footnotesize\ttfamily\bfseries,
            every two node part/.style={font=\ttfamily\footnotesize}
         },
]
\Tree [.\node [anchor=west,xshift=-2cm] {DataType};
    [. ScalarType
        [.f32 ]
        [.\node {$\ldots$}; ]
    ]
    [.\node {
        Array
        \nodepart{two}
        size: Nat;\\
        elem: DataType
    }; ]
    [.\node {
        Tuple
        \nodepart{two}
        fst: DataType;\\
        snd: DataType
    }; ]
]
\end{scope}
\begin{scope}[
    grow'=right,
    yshift=-5cm,
    xshift=-2cm,
    edge from parent path={(\tikzparentnode.east) -- +(+17pt,0) |- (\tikzchildnode.west)},
    level distance=60pt
]
\Tree [.\node[rectangle split horizontal=false] {
                \emph{Nat}
            };
    [.\node {
        Literal
        \nodepart{two}
        value: Int
     }; ]
    [.\node {
        BinOp
        \nodepart{two}
        op: String;
        lhs: Nat;
        rhs: Nat
    }; ]
    [.\node {$\ldots$}; ]
]
\end{scope}
\end{tikzpicture}
\caption{Grammar of \Rise{} types and type-level values (above) and diagram of classes (below) representing them.}
\label{fig:rise-type-overview}
\end{figure}

\Cref{fig:rise-type-overview} shows the grammar and corresponding class hierarchies of \Rise{} types.
Type level values are organized into distinct and interconnected classes following the formalization given by the grammar and the formal typing rules (not shown in this paper, but described in~\cite{DBLP:conf/cgo/HagedornSSGD18,DBLP:journals/corr/abs-1710-08332}).
\texttt{DataType}s are kept distinct from general types and represent types that can be stored in memory.
As we compile to GPUs, this design carefully avoids the challenge of storing functions (or closures) in memory, enforcing an important invariant of our IR.

\texttt{Array} types track their length with the member \texttt{size} that is a natural number and forms a distinct category of expressions represented by the \texttt{Nat} class and subclasses.

Finally, \cref{lst:rise-primitives} shows some important high-level data-parallel primitives modeled as functions.
We use a functional notation for their types, i.e., every argument is separated by $\rightarrow$.
For example, \texttt{map} expects 5 arguments, 3 at the type-level:
a natural number $n$, 2 data types $s$ and $t$;
and 2 ordinary arguments:
a function and an array to whose elements the function is applied to produce the output array.
Parameters in curly braces are implicit and automatically inferred.

\begin{listing}[t]
\begin{RISElisting}
map: {n: Nat} -> {s: DataType} -> {t: DataType} ->
  (s -> t) -> Array[n, s] -> Array[n, t]
reduce: {n: Nat} -> {t: DataType} ->
  (t -> t -> t) -> t -> Array[n, t] -> t
zip: {n: Nat} -> {s: DataType} -> {t: DataType} ->
  Array[n, s] -> Array[n, t] -> Array[n, Tuple[s, t]]
\end{RISElisting}
\caption{Selection of \Rise{} functional data-parallel primitives}
\label{lst:rise-primitives}
\end{listing}

\subsection[Optimizing RISE via rewriting]{Optimizing \Rise{} via rewriting}
Optimizing is the primary focus of \Rise{}, and it is clearly separated from compiling the functional abstractions away.
Translation to imperative constructs happens in later stages.

The rewrite rules used in the \Shine{} compiler follow closely the rules introduced by \Lift{}, for example, three algorithmic rules for splitting and fusing \texttt{map} and \texttt{reduce}:
\begin{RISElisting}
    map(f) $\mapsto$ split(n) >> map(map(f)) >> join
    map(f) >> map(g) $\mapsto$ map(f >> g)
    map(f) >> reduce(fun(acc, y => acc $\oplus$ y), init)
        $\mapsto$ reduceSeq(fun(acc, y => acc $\oplus$ f(y)), init)
\end{RISElisting}

These rules encode algorithmic optimization \emph{choices} that when applied to an expression directly encode that choice in the rewritten expression.
This becomes even more clear for rules translating a high-level \Rise{} primitive into a low-level implementation-specific primitive.
For example, the following rules provide different options to implement the high-level \texttt{map} primitive in the OpenCL programming model:
\begin{RISElisting}
    map $\mapsto$ mapSeq                 map $\mapsto$ mapGlobal
    map $\mapsto$ mapWorkGroup           map $\mapsto$ mapLocal
\end{RISElisting}

Applying one of these rules to a \texttt{map} primitive encodes the explicit choice that this primitive should be implemented either sequentially or in one of three different parallel ways, each exploiting different aspects of the GPU.

Rewrite rules transforming \Rise{} are expressed in \Elevate{}~\cite{DBLP:journals/pacmpl/HagedornLKQGS20} a language for composing rewrite rules into larger \emph{optimization strategies}.
In this design, rewrite rules are functions taking a \Rise{} expression as input and returning the transformed expression or a failure state if the rewrite rule could be applied.
An \Elevate{} optimization strategy that describes a particular composition of rewrite rules corresponds to a \emph{schedule} in systems such as TVM or Halide describing precisely how the high-level \Rise{} program should be optimized and implemented.

\subsection{Optimizing Matrix-Vector-Multiplication}
There are many choices to be made when deciding how to optimize the matrix-vector-multiplication example from \cref{fig:rise-mv}.
A key idea of optimizing by rewriting in the \Shine{} compiler is to expose these choices rather than make them internally.

\Cref{lst:elevate-strategy} shows one possible (and for illustration purposes simple) way to optimize.
This \Elevate{} optimization strategy describes a sequence of rewrite steps that transforms the high-level description of matrix-vector-multiplication into the low-level version shown in \cref{lst:mv-opt-rise}.
The outermost \lstinline!map! is split into two maps (using the \lstinline!splitJoinMap! rule) that are then lowered into two specific versions of map reflecting the OpenCL parallelism model.
The \lstinline!reduce! and \lstinline!map! primitives describing the dot product computation are fused into a single reduction (using the \lstinline!fuseReduceMap! rule).
This reduction can no longer be parallelized (as the operator is not associative), but we avoid the need for an intermediate buffer and two separate loops.

\begin{listing}[t]
\begin{RISElisting}
    splitJoinMap    `@` outermost(isMap)    `;`
    toMapWorkGroup  `@` outermost(isMap)    `;`
    toMapLocal      `@` outermost(isMap)    `;`
    fuseReduceMap   `@` every(isReduce)     `;`
    toReduceSeq     `@` every(isReduce)
\end{RISElisting}
\caption{
    One possible way to optimize Matrix-Vector-Multiplication encoded as an \Elevate{} optimization strategy.
}
\label{lst:elevate-strategy}
\end{listing}

\begin{listing}[t]
\begin{RISElisting}[numbers=left,xleftmargin=10pt]
def mvOpt = depFun((n: Nat, m: Nat) =>
  fun(M: Array[n, Array[m, f32]] =>
    fun(x: Array[m, f32] =>
      M |> split($s$) |> mapWorkGroup(fun(rows =>
        rows |> mapLocal(fun(row =>
          zip(row)(x) |>
            reduceSeq(Private)(fun(acc, ax =>
                acc + (fst(ax) * snd(ax))))(0.0f) ))
        )) |> join
    )) )
\end{RISElisting}
\caption{
    Optimized Matrix-Vector-Multiplication in \Rise{}.
}
\label{lst:mv-opt-rise}
\end{listing}

\subsection*{Summary}
In this section, we have discussed the design of the \Rise{} language that is used in the \Shine{} compiler for explicitly encoding optimization and implementation choices.
This clear purpose heavily influences the design of the language:
optimizations are expressed as compositions of rewrite rules, allowing them to be easily extended and externally controlled;
to enable rewriting, the IR is functional and free of side effects to be referentially transparent (i.e., to allow any expression to be replaced by an equivalent expression).

Deliberately, \Rise{} does not have imperative features or concepts, such as assignments into memory, or \texttt{for} loops.
This still leaves a large gap to close between the optimized \Rise{} program and the targeted imperative programming models, such as C, OpenMP, or OpenCL.
To translate the functional to imperative code, we introduce a separate dedicated language called \DPIA{} that does provide functional and imperative concepts and is, therefore, suited to serve for the further code generation process.
While existing functional solutions, such as \Lift{}, have \Rise{}-like languages, are missing a dedicated IR for code generation, such as \DPIA{}, and, therefore, combine optimizing and code generation in one step, complicating both steps and the overall compiler design.

\section{Type Systems for Formalizing Invariants~and~Assumptions}
\label{sec:type-system}
Each compiler IR has invariants and assumptions that must hold for optimizations and translations to work as expected.
For example, a polyhedral-specific IR must ensure that array indices are affine for polyhedral optimizations to be valid.
In this section, we will discuss how a language-oriented view helps to formalize such invariants and assumptions in the type system of the language used as the IR.

We already discussed in \cref{sec:optimizing-vs-lowering} that \Rise{} separates data types whose values are stored in memory from function types, enforcing an important invariant and ruling out programs that we are unable to compile into efficient GPU code.
We also formulated the clear goal, that all optimization and implementation decisions have to be encoded directly into the \Rise{} program.
This is a crucial assumption we make in the further compilation process.
But how do we enforce it?

It is easy to check that implementations have been chosen for high-level primitives, for example, it is easy to check that all \texttt{map} primitives have been replaced with their implementation-specific counterparts (e.g., \texttt{mapSeq} or \texttt{mapGlobal}).
But there is a crucial hidden choice that cannot be overlooked: memory.
Let's consider for example the following \Rise{} program:
\begin{listing}[h]
\begin{RISElisting}
depFun((n: Nat, m: Nat) =>
  fun(M: Array[n, Array[m, f32]] =>
    M |> mapWorkGroup(fun(row =>
      row |> mapLocal(f) |> mapLocal(g)) )))
\end{RISElisting}
\caption{
    \Rise{} program without choice of temporary memory.
}
\label{fig:rise-missing-memory}
\end{listing}

\noindent
For every element in every row of the matrix \texttt{M}, first the function \texttt{f} is applied and then the function \texttt{g}.
The parallelization strategy is clearly described here and also the fact that \texttt{f} and \texttt{g} will be computed in two separate steps rather than a single one.
But this leaves the choice of where to allocate the intermediate memory.
On a GPU, there are different address spaces (e.g., local/shared vs global memory) with vastly different size and performance characteristics.
In \Lift{}, this choice is already modeled via specific primitives such as \texttt{toLocal} and \texttt{toGlobal}~\cite{DBLP:conf/cgo/SteuwerRD17}, but making this choice explicitly is not enforced.
In the following, we introduce the \DPIA{} language, and its type system that formalizes the intuition that everywhere where we need to write to memory the address space choice has been made.
Furthermore, \DPIA{} combines functional and imperative aspects to facilitate the translation of the optimized functional program to imperative high-performance code.

\subsection[DPIA phrases, types and primitives]{\DPIA{} phrases, types, and primitives}

\DPIA{} is a variation of \citeauthor{Reynolds1997}'s idealized ALGOL~\cite{Reynolds1997,DBLP:journals/corr/abs-1710-08332} and aims to integrate functional and imperative concepts in a single language.
For this, we separate our program \emph{phrases} into three categories:
(functional) \emph{expressions}, (imperative) \emph{commands} (like C statements), and (imperative) \emph{acceptors}.
One can think of an acceptor as a pointer that we can manipulate and ultimately use for writing to a memory location.


\begin{figure}
\begin{lstlisting}[basicstyle=\ttfamily\footnotesize, commentstyle=\itshape\color{black!60}, mathescape,language=c,escapeinside={!:}{:!},]
$P$ := x | 0.0f |           !:\Comment variables and literals:!
    fun(x => $P$) |           !:\Comment function abstraction:!
    $P$ |> $P$ | $P$($P$) |   !:\Comment function application:!
    depFun(x: $K'$ => $P$) |   !:\Comment dependent fun. abstraction:!
    $P$($N$) | $P$($DT$) | $P$($A$) | !:\Comment dependent fun. application:!
    mapWorkGroup|reduceSeq|...  !:\Comment functional primitives:!
    $P$=$P$|(;)|new|parForWorkGroup|... !:\Comment imperative primitives:!

$T'$ := t | $T'\rightarrow{}T'$ | (x:$K$)$\,\rightarrow{}T'$| !:\Comment type var. \& function types:!
     $T'$ $\times$ $T'$ | !:\Comment phrase pair type:!
     Exp[$DT$, $RW$] | !:\Comment expression type:!
     Acc[$DT$] | !:\Comment acceptor type:!
     Comm !:\Comment command type:!
$RW$ :=  Rd | Wr     !:\Comment read-write annotations:!
$K'$ := $K$ | ReadWrite !:\Comment kinds:~ read-write \& RISE kinds:!
\end{lstlisting}
\caption{Grammar of \DPIA{} phrases, types, and type-level values.}
\label{fig:dpia-type-overview}
\end{figure}


\Cref{fig:dpia-type-overview} shows the grammar of \DPIA{} phrases and types.
The language looks similar to \Rise{} and in fact uses \Rise{}'s functional language constructs and data types.
New are the imperative constructs, such as assignment ($P$\texttt{=}$P$).
The type system reflects the separation into expressions (\lstinline!Exp[$DT$,$RW$]!), acceptors (\lstinline!Acc[$DT$]!), and commands (\lstinline!Comm!).
(Dependent) functions and pairs can freely combine phrases of all categories.
The expression type has a \lstinline!ReadWrite! annotation used for checking that an input program has encoded all choices about memory address spaces explicitly, as we will discuss in \cref{sec:read-write-annotations}.

\Cref{lst:dpia-functional-primitives} and \ref{lst:dpia-imperative-primitives} show a selection of \DPIA{} primitives.
In \RISE{}, we introduced primitives with a function type.
In contrast, in \DPIA{} primitives are fully applied values.
For example, in \RISE{} \lstinline!mapSeq(f)! represents the function application of \lstinline!mapSeq! to \lstinline!f!.
The equivalent program in \DPIA{}, must use a lambda: \lstinline!fun(x => mapSeq(f, x))!.
Insisting on fully applied primitives simplifies the translation and code generation process, while allowing partially applied primitives (such as \lstinline!map(f)!) simplifies the specification of rewrite rules.
With two separate IRs, we can choose the most convenient style for each task.

The functional primitives in \cref{lst:dpia-functional-primitives} only contain low-level variations of the high-level \lstinline!map! and \lstinline!reduce! primitives, as for programs at this stage, all implementation choices must have been made.
Some functional primitives, such as \lstinline!zip!, \lstinline!join!, and \lstinline!split! will always be fused into successive primitives and, therefore, remain unchanged.
There are also low-level primitives such as indexing into arrays (\lstinline!idx!) and \lstinline!toMem! which we will discuss in the next section.

\Cref{lst:dpia-imperative-primitives} shows a selection of \DPIA{}'s imperative primitives and their types.
These are commands, such as writing to memory via an \lstinline!assign!ment, \lstinline!seq!uencing commands, memory allocation (\lstinline!new!), and sequential and parallel \lstinline!for! loops.
\lstinline[basicstyle=\ttfamily\footnotesize\bfseries]!zipAcc1/2! and \lstinline!joinAcc! represent computations over acceptors, for example, treating a one-dimensional array as a two-dimensional one (\lstinline!joinAcc!).
Eventually, these computations will correspond to pointer manipulations in the generated code.



\begin{listing}[t]
\begin{RISElisting}
mapLocal(n: Nat, s: DataType, t: DataType,
  f: Exp[s,Rd] -> Exp[t,Wr],
  x: Exp[Array[n,s],Rd]): Exp[Array[n,t],Wr]
reduceSeq:(n: Nat, a: AddrSp, s: DataType, t: DataType,
  f: Exp[t,Rd] -> Exp[s,Rd] -> Exp[t,Wr],
  init: Exp[t,Wr], x: Exp[Array[n,s],Rd]): Exp[t,Rd]
zip:(n: Nat, s: DataType, t: DataType, w: ReadWrite,
  lhs: Exp[Array[n,s],w],
  rhs: Exp[Array[n,t],w]): Exp[Array[n,Tuple[s,t]],w]
join:(n: Nat, m: Nat, t: DataType, w: ReadWrite,
  x: Exp[Array[n,Array[m,t]],w]): Exp[Array[n*m,t],w]
split:(n: Nat, m: Nat, t: DataType, w: ReadWrite,
  x: Exp[Array[n*m,t],w]): Exp[Array[m,Array[n,t]],w]

toMem:(a: AddrSp, t: DataType, x: Exp[t,Wr]): Exp[t,Rd]
idx:(n: Nat, t: DataType,
  i: Exp[Idx[n],Rd], arr: Exp[Array[n,t],Rd]): Exp[t,Rd]
\end{RISElisting}
\caption{Selection of \DPIA{} functional primitives}
\label{lst:dpia-functional-primitives}
\end{listing}

\begin{listing}[t]
\begin{RISElisting}
assign:(t: DataType, lhs: Acc[t], rhs: Expr[t, Rd]): Comm
seq:(c1: Comm, c2: Comm): Comm
new:(a: AddrSp, t: DataType,
  body: (Exp[t,Rd] x Acc[t]) -> Comm): Comm
for:(n: Nat, body: Exp[Idx[n],Rd] -> Comm): Comm
parForLocal(n: Nat, t: DataType,
  out: Acc[Array[n,t]],
  body: Exp[Idx[n],Rd] -> Acc[t] -> Comm): Comm
    
zipAcc1:(n: Nat, s: DataType, t: DataType,
  array: Acc[Array[n,Tuple[s,t]]]): Acc[Array[n,s]]
zipAcc2:(n: Nat, s: DataType, t: DataType,
  array: Acc[Array[n,Tuple[s,t]]]): Acc[Array[n,t]]
joinAcc:(n: Nat, m : Nat, t : DataType,
  array: Acc[Array[n*m,t]]): Acc[Array[n,Array[m,t]]]
\end{RISElisting}
\caption{Selection of \DPIA{} imperative primitives}
\label{lst:dpia-imperative-primitives}
\end{listing}

\subsection{Enforcing Presence of Necessary Address Space Choices}
\label{sec:read-write-annotations}
In \cref{fig:rise-missing-memory} we introduced an example of a \Rise{} program where a necessary choice to clarify the address space of a temporary memory buffer was missing.
If we attempt to compile this program, we have to make a decision on behalf of the user.
We want to avoid such situations, and \DPIA{}'s type system formalizes this:
when translating a program from \Rise{} to \DPIA{} the type system checks the consistency of all \lstinline!ReadWrite! annotations and rejects programs with inconsistent annotations.
%
Let's revisit the example from \cref{fig:rise-missing-memory}, after it has been translated to \DPIA{}:
\begin{RISElisting}
depFun((n: Nat, m: Nat) =>
  fun(M: Exp[Array[n, Array[m, f32]], Rd] =>
    mapWorkGroup(n, Array[m, f32], Array[m, f32],
      fun(row => mapLocal(m, f32, f32, g,
                    mapLocal(m, f32, f32, f, row))),
      M )))
\end{RISElisting}
The translation process from low-level \Rise{} to functional \DPIA{} is straightforward, as there exists a corresponding \DPIA{} primitive for each low-level primitive in \Rise{}.

The program should not type check in \DPIA{} as a choice of address space is missing.
Let us understand why by looking at the crucial part of the program:
%
\begin{RISElisting}[escapeinside=~~]
mapLocal(m, f32, f32, g,
  // expected:               Exp[Array[m, f32], Rd] ~\cross{}~
  mapLocal(m, f32, f, row) : Exp[Array[m, f32], Wr])
\end{RISElisting}

As indicated by the comment, we can see that the type of the second \lstinline!mapLocal! does not match the type expected by the first \lstinline!mapLocal!.
The precise type of \lstinline!mapLocal! is given in \cref{lst:dpia-functional-primitives}.
Clearly, the \lstinline!ReadWrite! annotations do not match:
we expect a value that we can read from memory (as indicated by \lstinline!Rd!), but instead we have been given a value that first has to be written to memory (as indicated by \lstinline!Wr!) before we can read it.

The \lstinline!ReadWrite! annotations are introduced in the translation process from \Rise{} to \DPIA{} and the functional primitives in \cref{lst:dpia-functional-primitives} are explicitly annotated with them.
Some primitives, such as \lstinline!zip!, are polymorphic over \lstinline!ReadWrite! annotations, i.e., their annotation depends on the context they are used in.
For these primitives, a simple inference is performed.

There exists only one primitive to convert an expression that needs to be written to memory (i.e., with \lstinline!Wr!) to an expression from which we can read (i.e., with \lstinline!Rd!):
\lstinline!toMem! which expects the address space to write into as an argument.

Therefore, to fix our program, we need to inject a \lstinline!toMem! in between the two \lstinline!mapLocal!s.
This choice has to be encoded in the \Rise{} program, e.g., via an appropriate rewrite rule, before we translate to 
\DPIA{}.
Once we are in \DPIA{}, we can only check that address space annotations are consistent.

After adding \lstinline!toMem! in the \Rise{} program and translating to \DPIA{} we see that the type checking succeeds:
\begin{RISElisting}[escapeinside=~~]
mapLocal(m, f32, f32, g,
 // expected: Exp[Array[m, f32], Rd] ~\checkmark{}~
 toMem(Private, Array[m, f32],
  // expected:               Exp[Array[m, f32], Wr] ~\checkmark{}~
  mapLocal(m, f32, f, row) : Exp[Array[m, f32], Wr]
 )          : Exp[Array[m, f32], Rd] )
\end{RISElisting}

It is important to note that \lstinline!toMem! only accepts arguments that are not yet written to memory, as indicated by the \lstinline!Wr! annotation on its input.
Otherwise, we would have to make a decision on \emph{how} to copy the value from one memory location to another.
But there are many ways to copy, for example, an array: sequentially, vectorized, or in parallel using threads?
As we refuse to make such choices in \DPIA{}, we reject such programs and insist that the choice on how to perform the copy must be encoded in \Rise{} explicitly.

\subsection{Translating Functional into Imperative Programs}

\begin{listing}[t]
\begin{RISElisting}[escapeinside=~~,mathescape=true]
def accT(expr: Phrase[Exp[d,Wr]],
         output: Phrase[Acc[t]]
        ): Phrase[Comm] = expr match {
 case mapWorkGroup(n, s, t, f, arr) =>
  conT(arr, fun(arrT =>
    parForWorkGroup(n, t, output, fun((i, o) =>
      accT(f(idx(n, s, i, arrT)), o) ))))
 case mapSeq(n, s, t, f, arr) =>
  conT(arr, fun(arrT =>
    for(n, fun(i =>
      accT(f(idx(n,s,i,arrT)), idxAcc(n,t,i,output)) ))))
 case zip(n, s, t, Wr, lhs, rhs) =>
  accT(lhs, zipAcc1(n, s, t, output)) ;
  accT(rhs, zipAcc2(n, s, t, output))
 case join(n, m, t, Wr, arr) =>
  accT(arr, joinAcc(n, m, t, output))
 case ...
}
\end{RISElisting}
\caption{Definition of acceptor translation}
\label{lst:dpia-acceptor-translation}
\end{listing}

\begin{listing}[t]
\begin{RISElisting}[escapeinside=~~,mathescape=true]
def conT(expr: Phrase[Exp[t,Rd]],
         continuation: Phrase[Exp[t,Rd]] -> Phrase[Comm]
        ): Phrase[Comm] = expr match {
 case toMem(a, t, e) =>
  new(a, t, fun((tmpAcc, tmpExp) =>
    accT(e, tmpAcc) ; continuation(tmpExp) ))
 case reduceSeq(n, a, t, s, f, init, arr) =>
  conT(arr, fun(arrT =>
    new(a, t, fun((accumAcc, accumExp) =>
      accT(init, accumAcc) ;
      for(n, fun(i =>
        accT(f(accumExp, idx(n, s, i, arrT)),
             accumAcc) )) ;
      conT(accumExp, continuation) )) ))
 case zip(n, s, t, Rd, lhs, rhs) =>
  conT(lhs, fun(lhsT =>
    conT(rhs, fun(rhsT =>
      continuation(zip(n, s, t, Rd, lhsT, rhsT))))))
 case join(n, m, t, Rd, x) =>
  conT(x, fun(xT => continuation(join(n, m, t, Rd, xT))))
 case split(n, m, t, Rd, x) =>
  conT(x, fun(xT => continuation(split(n, m, t, Rd, xT))))
 case fst(s, t, Rd, x) =>
  conT(x, fun(xT => continuation(fst(s, t, Rd, xT))))
 case snd(s, t, Rd, x) =>
  conT(x, fun(xT => continuation(snd(s, t, Rd, xT))))
 case ...
}
\end{RISElisting}
\caption{Definition of continuation translation}
\label{lst:dpia-continuation-translation}
\end{listing}

Now that we are confident that all implementation decisions are encoded in the functional \DPIA{} program, we discuss how it is translated into a program using \DPIA{}'s imperative features.
For example, we want to translate a primitive such as \lstinline!reduceSeq! into a \lstinline!for! loop.

This translation process is performed by two intertwined translation functions, shown in \cref{lst:dpia-acceptor-translation} and \ref{lst:dpia-continuation-translation}.
The \emph{acceptor translation} \lstinline!accT! takes a functional expression that needs to be written to memory (i.e., with \lstinline!Wr!) plus an acceptor representing the memory to write into.
The \emph{continuation translation} \lstinline!conT! takes expressions that we can read from (i.e., with \lstinline!Rd!) plus a \emph{continuation} function that knows how to continue the translation once we have translated the current expression.
Both translations return an imperative command.

To start the overall translation process, we allocate \emph{output} memory according to the data type of the value that the functional program computes.
When compiling to OpenCL, the results of a kernel must reside in global memory so that we have no choice of address space to make here.
With the given \emph{output}, we invoke the \lstinline!accT! translation.
This corresponds to the intuition of translating a program that computes a value functionally and afterwards writes it to memory.

For the matrix-vector-multiplication translated from \Rise{} (\cref{lst:mv-opt-rise}) to \DPIA{} we invoke \lstinline!accT! with an output array:
\begin{RISElisting}
mvOptImp = accT(mvOpt, $\textit{output}$ : Acc[Array[n,f32]])
\end{RISElisting}

Following the intuition of the acceptor translation, we traverse the functional program backwards, starting with the last primitive; the primitive that writes into the output.
%
In case of the matrix-vector multiplication, the last primitive is \lstinline!join! with \lstinline!mapWorkGroup! as its argument.
\lstinline!join! changes our view from a two-dimensional into a one-dimensional array.
The acceptor translation indicates that we call \lstinline!accT! with \lstinline!join!'s argument (here \lstinline!mapWorkGroup!) and wrap the output in a \lstinline!joinAcc!:
\begin{RISElisting}
accT(join(mapWorkGroup(f, arr)), output)
  = accT(mapWorkGroup(f, arr), joinAcc(output))
\end{RISElisting}

The \lstinline!joinAcc! primitive has the reverse effect on an acceptor to that of \lstinline!join! on an expression:
we transform our view on the output memory, instead of a flat one-dimensional array we treat it as two-dimensional.
Then we translate \lstinline!mapWorkGroup!:
\begin{RISElisting}
accT(mapWorkGroup(f, arr), joinAcc(output))
  = conT(arr, fun(arrT =>
      parForWorkGroup(n/s, joinAcc(output), fun((i, o) =>
          accT(f(idx(i, arrT)), o) ))))
\end{RISElisting}

In the imperative program, the code computing the input of \lstinline!mapWorkGroup! must appear before the code for \lstinline!mapWorkGroup! itself, therefore, we must continue to translate the input first by using the \emph{continuation translation}.
This is the right choice, as we must be able to read from this expression based on the type of \lstinline!mapWorkGroup!.
To describe how the translation continues, we pass along a \emph{continuation} function that will be called once we have translated the input.
This function will be given the translated input array (\lstinline!arrT!) and inside we describe the translation of \lstinline!mapWorkGroup! itself into a \lstinline!parForWorkGroup!.
This parallel for loop is parameterized with the length of the array, it is mapping over; the output it is writing into; and a function representing the loop body.
The loop body has access to the loop index \lstinline!i! as well as an acceptor \lstinline!o! that guarantees that each iteration writes to a distinct location in the output array.
In the loop body, we invoke the acceptor translation of the function \lstinline!f! applied to a single value of the (translated) input array by indexing into it and passing the acceptor to write into along.


To translate the full program, we continue the translation process by calling the \lstinline!accT! and \lstinline!conT! functions as indicated in \cref{lst:dpia-acceptor-translation} and \ref{lst:dpia-continuation-translation} for each primitive, until we have traversed the entire functional program.

The result of the translation process is the imperative \DPIA{} program shown in \cref{lst:dpia-matrix-vector-mult}, from which we generate the OpenCL code shown in \cref{lst:opencl-matrix-vector-mult}.
This final code generation step is small, as the structure of the imperative \DPIA{} program directly corresponds to the OpenCL program.
For example, \lstinline!parForWorkGroup! and \lstinline!parForLocal! correspond to the two outer loops in line 6 and 8 in the OpenCL code.
The remaining gap is to resolve some functional expressions into multi-dimensional array indices.
This process is similar to the \emph{views} in \Lift{}~\cite{DBLP:conf/cgo/SteuwerRD17} and involves simplifying arithmetic expressions to generate concise indices.
The only conceptual difference to \Lift{} is that in \DPIA{} index computations also happen on acceptors to manipulate the location where we write into memory.

\begin{listing}[t]
\begin{RISElisting}[numbers=left,xleftmargin=10pt]
depFun((n: Nat, m: Nat) =>
  fun(M:Array[n,Array[m,f32]] => fun(x:Array[m,f32] =>
parForWorkGroup(n/$s$, Array[m,f32], joinAcc($\textit{output}$),
 fun((wgId, wgOut) =>
  parForLocal($s$,f32, wgOut, fun((lId, lOut) => 
   new(Private, f32, fun((accumAcc, accumExp) =>
    accumAcc = 0.0f;
    for(m, fun(i =>
     accumAcc = accumExp +
      fst(idx(i,zip(idx(lId,idx(wgId,split($s$,M))),x)))
    * snd(idx(i,zip(idx(lId,idx(wgId,split($s$,M))),x)))
    )) ; lOut = accumExp )))))))))
\end{RISElisting}
\caption{MV-Mult. after translation into imperative \DPIA{}}
\label{lst:dpia-matrix-vector-mult}
\end{listing}

\begin{listing}
\begin{minted}{c}
__kernel
  void mvOptKernel(global float* restrict output,
                   int n, int m, int s,
                   const global float* restrict M,
                   const global float* restrict x) {
for (int wgId = get_group_id(0);   // parForWorkGroup
     wgId < n/s; wgId += get_num_groups(0)) {
 for (int lId = get_localId(0);    //     parForLocal
      lId < s; lId += get_local_size(0)) {
   float accum;                             //    new
   accum = 0.0f;                            // assign
   for (int i = 0; i < m; i += i) {         //    for
     accum += (M[(i+lId*m)+(m*s*wgId)] * x[i]); }
   output[lId + (s * wgId)] = accum; }}}    // assign
\end{minted}
\caption{OpenCL code generated for MV-Mult.}
\label{lst:opencl-matrix-vector-mult}
\end{listing}

\subsection*{Summary}
In this section, we discussed the \DPIA{} language that is used as an IR in the \Shine{} compiler to translate functional to imperative programs.
We have seen, how our language-oriented compiler design enables the formalization of important invariants and assumptions in the type system and prevents ad-hoc implementation choices in the compilation process.
Instead, all implementation choices are pushed upwards and must have been encoded explicitly beforehand.

\section{Compiler Extensibility at Multiple Levels}
\label{sec:extensibility}

Extensibility is a crucial consideration in compiler design.
By having a slim core language, both \Rise{} and \DPIA{} are easy to extend by adding new primitives.
A crucial advantage over purely functional IRs, such as \Lift{}, is that we can extend \DPIA{} with new imperative primitives, directly capturing code patterns or interfaces in the target programming model.

We illustrate the extensibility by adding support for iterative computations.
\Lift{} proposes an \lstinline!iterate! primitive, that repeatedly applies a function to an array, passing the output of one iteration as the input of the next.
This can be used to describe tree-based reductions and similar iterative algorithms~\cite{DBLP:conf/icfp/SteuwerFLD15}.

One efficient way to implement such computations in a C-like programming language is to use double buffering by swapping pointers to two buffers after each iteration.
But how would we represent this in \DPIA{}, as there is no support for dealing with pointers directly?
To solve this, we add two primitives to \DPIA{}: the functional \lstinline!iterate! and the imperative \lstinline!newDoubleBuffer!.

To start, let's look at the C code we would like to generate for an iterative computation implemented using double buffering, as shown in \cref{lst:double-buff-code}.
First we need to allocate temporary buffers and then initialize two pointers \lstinline!in_ptr! and \lstinline!out_ptr! who are pointing to the input and output of each iteration.
The \lstinline!for! loop iterates for a fixed number of \lstinline!k! iterations and performs the iterative computation in line 5.
Afterwards, we need to swap the pointers to prepare for the next iteration.
Because we want to avoid the need for unnecessary copies, in the first iteration \lstinline!in_ptr! points to the input of the overall computation and in the last \lstinline!out_ptr! to the output.
To achieve this, we use a \lstinline!flag! indicating how we have to swap and an \lstinline!if! branch to check if the last iteration has been reached.

\begin{listing}
\begin{minted}{c}
float buffer1[size]; float buffer2[size];
float* in_ptr = input; float* out_ptr = buffer1;
unsigned char flag = 1;
for (int i = 0; i < k; i += 1) {
   body(in_ptr, out_ptr);             // body
   if (i <= (k-1)) {                  // swap
      in_ptr = flag ? buffer1 : buffer2;
     out_ptr = flag ? buffer2 : buffer1;
     flag = flag ^ 1;
   } else {                           // done
      in_ptr = flag ? buffer1 : buffer2;
     out_ptr = output; } }
\end{minted}
\caption{C code expressing double buffering.}
\label{lst:double-buff-code}
\end{listing}

When adding new primitives, we need to model these implementation aspects in \DPIA{}.
For double buffering, we want to avoid exposing pointers directly, and thus we hide the pointer manipulation behind two commands that we expose in \DPIA{}.
\Cref{lst:extension} shows the added \lstinline!newDoubleBuffer! primitive.
It expects multiple arguments:
the data types and sizes of the input, output, and temporary buffers,
the input array, the output array, and a \lstinline!body! function that has access to the double buffer.
The \lstinline!body! expects four arguments:
an acceptor and expression to write/read into/from the current active buffer, and two commands corresponding to the \emph{swap} and \emph{done} code parts from \cref{lst:double-buff-code}.
The \lstinline!body! may use \emph{swap} and \emph{done} freely.
In the final code generation process, these opaque commands are replaced with the code shown in the listing.


\cref{lst:accT-extension} shows the translation of the functional \lstinline!iterate! primitive (List.~\ref{lst:extension}) to imperative \DPIA{}.
The \lstinline!iterate! primitive is translated into a combination of \lstinline!newDoubleBuffer!, to allocate a double buffer, and \lstinline!for!, to express the iterative computation.
Specifically, the \lstinline!swap! and \lstinline!done! commands are used inside the \lstinline!if!, directly corresponding to the C code in \cref{lst:double-buff-code}.

\begin{listing}
\begin{RISElisting}
newDoubleBuffer(t: DataType, n: Nat, m: Nat, k: Nat,
  input: Exp[Array[m,t],Rd], output: Acc[Array[k,t]],
  body: (Acc[Array[n,t]$\,$x$\,$Exp[Array[n,t],Rd]$\,$x$\,$Comm$\,$x$\,$Comm)
       -> Comm): Comm
       
iterate(n: Nat, m: Nat, k: Nat, t: DataType,
  body: (l:Nat)->Exp[Array[l*n,t],Rd]->Exp[Array[l,t],Wr],
  arr: Exp[Array[(n^k)*m,t],Rd]): Exp[Array[m,t],Wr]
\end{RISElisting}
\caption{Extension to support iterative computations using double buffering.}
\label{lst:extension}
\end{listing}

\begin{listing}
\begin{RISElisting}
def accT(expr, output) = expr match {
 case ...
 case iterate(n, m, k, t, body, arr) =>
  conT(arr, fun(arrT => 
    newDoubleBuffer(t, (n^k)*m, m, (n^k)*m,
      arr, output, fun((tmpAcc, tmpExp, swap, done) =>
        for(k, fun(i =>
          body(n^(k-i-1)*m, tmpAcc, tmpExp) ;
          if (i < k) { swap } else { done } )))))
}
\end{RISElisting}
\caption{Extension of the acceptor translation}
\label{lst:accT-extension}
\end{listing}

\subsection*{Summary}
This case study of extensibility highlights two important points:
\emph{1)} the language-oriented compiler design with multiple single-purpose IRs allows adding abstractions at multiple levels, such as the high-level \lstinline!iterate! and the low-level \lstinline!newDoubleBuffer!;
\emph{2)} the abstractions stand on their own:
we did not design an imperative version of iterate, but instead designed a double buffer abstraction that can be used in other ways by using the exposed interface of \lstinline!swap! and \lstinline!done!.

\section{Evaluation}
\label{sec:evaluation}
We now want to evaluate our language-oriented compiler design.
We will do this in two ways:
\emph{1)} we will describe a qualitative comparison with the \Lift{} compiler highlighting the benefits of our approach over a compiler with similar design goals;
\emph{2)} we perform an experimental evaluation to confirm that the code generated by the \Shine{} compiler has the same runtime performance as code generated by \Lift{}.

\subsection{Qualitative comparison with \Lift{}}
From the user perspective, the \Lift{} compiler and \Shine{} have the same goal and \Rise{} has almost the same interface as the \Lift{} IR.
But their compiler design and implementations differ greatly.

Specifically, \Lift{} \emph{does not} follow a language-oriented compiler design.
It uses a single intermediate representation that lacks a formal type system.
For example, \Lift{} lacks the concept of a function type despite primitives being modeled as functions.
This makes specifying primitives cumbersome and the type inference implementation ad-hoc.
\Shine{} on the other hand, is implemented with two distinct languages that have designated purposes (optimizing vs. code generation).
This more modular design comes with significant benefits for engineering, but also conceptually.

First, optimizing by rewriting is performed in a purely functional language without compromising the language design for other tasks.
This is not the case in \Lift{}, where a single IR is used for both tasks and primitives are modelled not as functions, but as fully applied values, which makes the implementation of a rewrite system inconvenient.

Second, \DPIA{} -- the language for code generation -- formalizes and enforces the necessary assumptions of code generation.
This is not the case in \Lift{} where the compiler will fail with internal errors or is forced to make ad-hoc implementation choices when these have not been encoded in the input program.

Finally, extensibility in \Lift{} is limited to functional primitives, whereas in \Shine{} we can easily extend \DPIA{} with imperative primitives that provide a safe interface re-usable for compiling many high-level primitives.
The bottom up process of first establishing an imperative primitive in \DPIA{} and then thinking about how to compile a functional primitive to it, provides a principled pathway to incorporate implementation patterns and library interfaces of the low-level target language.

\subsection{Runtime performance compared to \Lift{}}
\begin{figure}
    \centering
    \includegraphics[width=\linewidth]{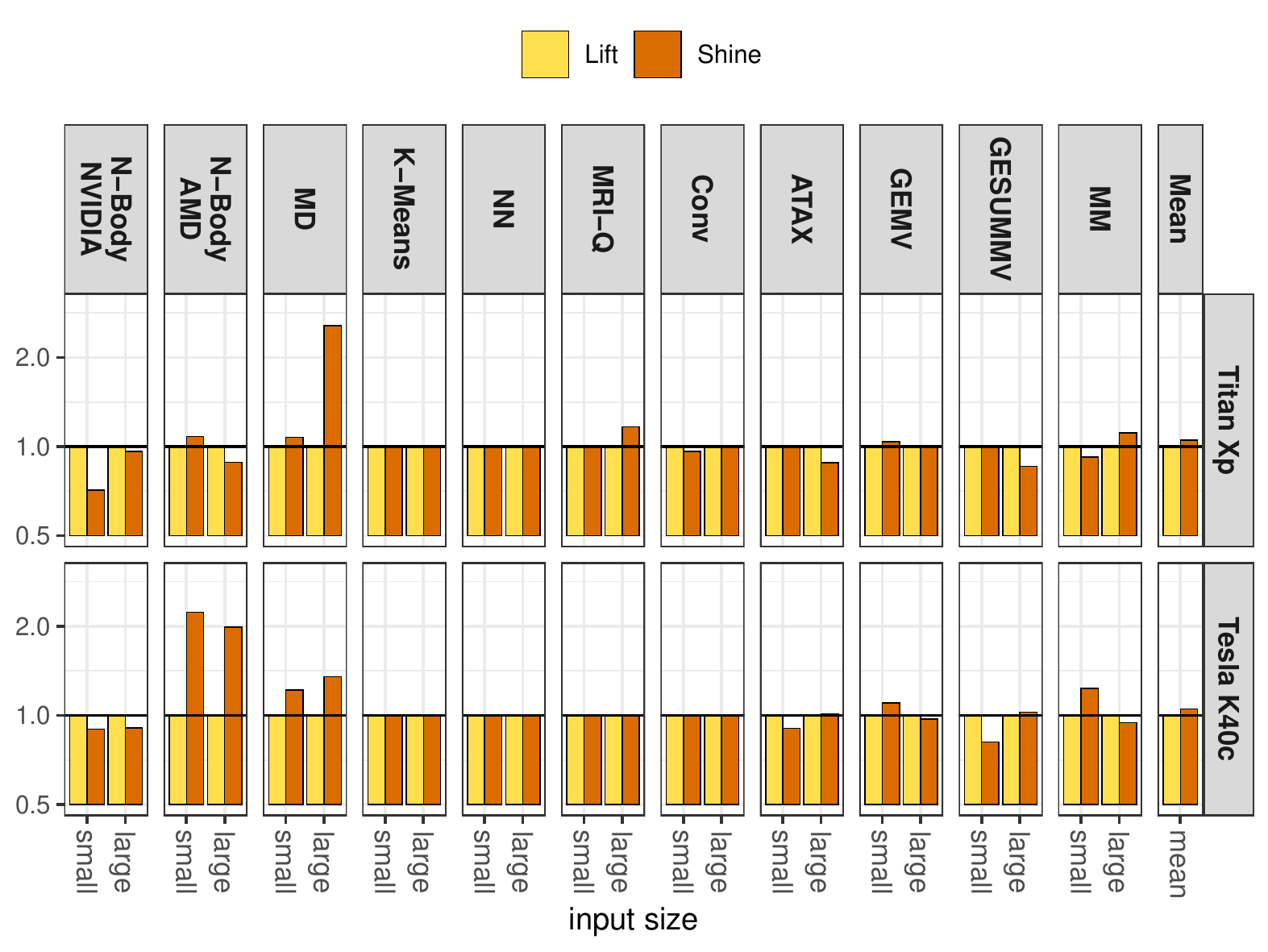}
    \caption{Relative runtime performance of the code generated by \Shine{} and \Lift{}.
    A higher bar indicates better performance.}
    \label{fig:runtime-benchmarks}
\end{figure}

We now evaluate the quality of the code generated by \Shine{} and \Lift{} using the benchmarks used in a previous \Lift{} paper~\cite{DBLP:conf/cgo/SteuwerRD17}.
We like to thank the authors for making their artifact publicly available.
\Cref{fig:runtime-benchmarks} shows the runtime performance on two Nvidia GPUs.
The benchmarks use the same optimizations in both compilers, even though \Lift{} makes some ad-hoc implementation choices, e.g., about memory allocation and vectorization, internally in the compiler rather than encoding them in the functional program as we do in \Shine{}.

We expect that the generated code has the same performance for both compilers, as (mostly) the same optimizations are applied.
For the mean across all benchmarks, this is indeed the case: the code generated by \Shine{} is slightly faster (about 5\%).
There are two significant outliers, for which \Shine{} generates significantly faster code:
\texttt{MD} and \texttt{N-Body AMD}.
Both benchmarks use vectorization and receive as input or produce as output arrays of \lstinline!float4! values.
While \Shine{} honors this interface and generates OpenCL kernels with \lstinline!float4! pointers, the \Lift{} compiler produces code with \lstinline!float! pointers and uses \texttt{vload}/\texttt{vstore} vector instructions to load/store values from these arrays.
In this case, this small code difference produces a significant performance gap.
The \texttt{NBody NVIDIA} benchmark shows a clear performance benefit for the \Lift{} generated code.
After inspection of the generated code, this can be explained by the lack of a memory re-use mechanism in \Shine{}, whereas the \Lift{} generated code uses less memory.
Reusing memory is currently not modelled in the more formal \DPIA{}, but could be added in the future.

\medskip
To summarize, \Shine{} shows the same performance as the \Lift{} compiler, while following a much better and more principled language-oriented compiler design.



\section{Related Work}
\label{sec:realted-work}
\paragraph*{Functional Compiler IRs}
We have extensively compared our work to the closest related work, \Lift{}~\cite{DBLP:conf/icfp/SteuwerFLD15,DBLP:conf/cgo/SteuwerRD17}, throughout this paper.
There exist several similar compilers with functional IRs, including Futhark~\cite{DBLP:conf/pldi/HenriksenSEHO17}, Accelerate~\cite{DBLP:conf/popl/ChakravartyKLMG11}, and Dex~\cite{DBLP:journals/pacmpl/PaszkeJDVRJRM21}.
Delite~\cite{DBLP:conf/cgo/BrownLRSSAO16} uses parallel patterns similar to \Rise{} to provide a framework for building DSL compilers.
\citeauthor{DBLP:conf/cc/LuckeSS21} presents an implementation of a \Rise{}-like language as a MLIR dialect~\cite{DBLP:conf/cc/LuckeSS21}.
The MLIR~\cite{DBLP:conf/cgo/LattnerAB0DPRSV21} dialect \texttt{linalg} cites \Lift{} as a direct influence.
Of course there exist other IRs that incorporate functional aspects, such as the functional graph-based Thorin~\cite{DBLP:conf/cgo/LeissaKH15} IR that provides explicit support for higher-order functions.
Many machine learning compilers, such as TensorFlow~\cite{DBLP:conf/osdi/AbadiBCCDDDGIIK16} and PyTorch~\cite{DBLP:journals/corr/abs-1912-01703}, follow a functional graph-based IR that is inspired by ideas from data-flow programming.


\paragraph*{Extensible Compilers}
\citeauthor{DBLP:conf/popl/RompfSABJLJOO13} present a technique based on staging for extensible compiler design~\cite{DBLP:conf/popl/RompfSABJLJOO13}.
Similarly, the Any DSL framework sees partial evaluation as the key to expressing compiler optimizations as easily extensible library code rather than internal IR transformations~\cite{DBLP:journals/pacmpl/LeissaBHPMSMS18}.
AnyDSL uses the functional Thorin IR mentioned above.
\citeauthor{DBLP:conf/cgo/KoehlerS21} argues for domain extensible compilers by extending the compiler with new image processing specific primitives and rewrite rules~\cite{DBLP:conf/cgo/KoehlerS21}.


\section{Conclusion}
\label{sec:conclusion}

In this paper, we present the \Shine{} compiler that follows a language-oriented compiler design.
The compiler is composed of multiple formal programming languages that are used as intermediate representations.

Each language has a clear purpose, separating optimizing clearly from the code generation process.
This simplifies both phases:
optimizing is expressed exclusively by rewriting the higher-level functional \Rise{} program encoding implementation and optimization choices;
code generation bridges from the functional to the imperative paradigm and preserves the encoded implementation choices.

\DPIA{}, the language for code generation, formalizes and checks the assumption that all implementation choices have been made before the code generation process starts.

By composing multiple languages, the compiler becomes easily extensible at various abstraction levels.
We demonstrated extending with an imperative primitive that captures a desirable low-level double buffering code pattern.

Finally, our evaluation showed the benefits of our compiler design over the closely related \Lift{} compiler while maintaining the same runtime code performance.

\newpage

\bibliographystyle{plainnat}
\bibliography{main}

\begin{thebibliography}{32}
\providecommand{\natexlab}[1]{#1}
\providecommand{\url}[1]{\texttt{#1}}
\expandafter\ifx\csname urlstyle\endcsname\relax
  \providecommand{\doi}[1]{doi: #1}\else
  \providecommand{\doi}{doi: \begingroup \urlstyle{rm}\Url}\fi

\bibitem[Abadi et~al.(2016)Abadi, Barham, Chen, Chen, Davis, Dean, Devin,
  Ghemawat, Irving, Isard, Kudlur, Levenberg, Monga, Moore, Murray, Steiner,
  Tucker, Vasudevan, Warden, Wicke, Yu, and
  Zheng]{DBLP:conf/osdi/AbadiBCCDDDGIIK16}
Mart{\'{\i}}n Abadi, Paul Barham, Jianmin Chen, Zhifeng Chen, Andy Davis,
  Jeffrey Dean, Matthieu Devin, Sanjay Ghemawat, Geoffrey Irving, Michael
  Isard, Manjunath Kudlur, Josh Levenberg, Rajat Monga, Sherry Moore,
  Derek~Gordon Murray, Benoit Steiner, Paul~A. Tucker, Vijay Vasudevan, Pete
  Warden, Martin Wicke, Yuan Yu, and Xiaoqiang Zheng.
\newblock Tensorflow: {A} system for large-scale machine learning.
\newblock In \emph{{OSDI}}, pages 265--283. {USENIX} Association, 2016.

\bibitem[Alpern et~al.(1988)Alpern, Wegman, and
  Zadeck]{DBLP:conf/popl/AlpernWZ88}
Bowen Alpern, Mark~N. Wegman, and F.~Kenneth Zadeck.
\newblock Detecting equality of variables in programs.
\newblock In \emph{{POPL}}, pages 1--11. {ACM} Press, 1988.

\bibitem[Appel(1998)]{DBLP:journals/sigplan/Appel88}
Andrew~W. Appel.
\newblock {SSA} is functional programming.
\newblock \emph{{ACM} {SIGPLAN} Notices}, 33\penalty0 (4):\penalty0 17--20,
  1998.

\bibitem[Atkey et~al.(2017)Atkey, Steuwer, Lindley, and
  Dubach]{DBLP:journals/corr/abs-1710-08332}
Robert Atkey, Michel Steuwer, Sam Lindley, and Christophe Dubach.
\newblock Strategy preserving compilation for parallel functional code.
\newblock \emph{CoRR}, abs/1710.08332, 2017.

\bibitem[Bravenboer et~al.(2008)Bravenboer, Kalleberg, Vermaas, and
  Visser]{DBLP:journals/scp/BravenboerKVV08}
Martin Bravenboer, Karl~Trygve Kalleberg, Rob Vermaas, and Eelco Visser.
\newblock Stratego/xt 0.17. {A} language and toolset for program
  transformation.
\newblock \emph{Sci. Comput. Program.}, 72\penalty0 (1-2):\penalty0 52--70,
  2008.

\bibitem[Brown et~al.(2016)Brown, Lee, Rompf, Sujeeth, Sa, Aberger, and
  Olukotun]{DBLP:conf/cgo/BrownLRSSAO16}
Kevin~J. Brown, HyoukJoong Lee, Tiark Rompf, Arvind~K. Sujeeth, Christopher~De
  Sa, Christopher~R. Aberger, and Kunle Olukotun.
\newblock Have abstraction and eat performance, too: optimized heterogeneous
  computing with parallel patterns.
\newblock In \emph{{CGO}}, pages 194--205. {ACM}, 2016.

\bibitem[Chakravarty et~al.(2011)Chakravarty, Keller, Lee, McDonell, and
  Grover]{DBLP:conf/popl/ChakravartyKLMG11}
Manuel M.~T. Chakravarty, Gabriele Keller, Sean Lee, Trevor~L. McDonell, and
  Vinod Grover.
\newblock Accelerating haskell array codes with multicore gpus.
\newblock In \emph{{DAMP}}, pages 3--14. {ACM}, 2011.

\bibitem[Chen et~al.(2018)Chen, Moreau, Jiang, Zheng, Yan, Shen, Cowan, Wang,
  Hu, Ceze, Guestrin, and Krishnamurthy]{DBLP:conf/osdi/ChenMJZYSCWHCGK18}
Tianqi Chen, Thierry Moreau, Ziheng Jiang, Lianmin Zheng, Eddie~Q. Yan, Haichen
  Shen, Meghan Cowan, Leyuan Wang, Yuwei Hu, Luis Ceze, Carlos Guestrin, and
  Arvind Krishnamurthy.
\newblock {TVM:} an automated end-to-end optimizing compiler for deep learning.
\newblock In \emph{{OSDI}}, pages 578--594. {USENIX} Association, 2018.

\bibitem[Hagedorn et~al.(2018)Hagedorn, Stoltzfus, Steuwer, Gorlatch, and
  Dubach]{DBLP:conf/cgo/HagedornSSGD18}
Bastian Hagedorn, Larisa Stoltzfus, Michel Steuwer, Sergei Gorlatch, and
  Christophe Dubach.
\newblock High performance stencil code generation with lift.
\newblock In \emph{{CGO}}, pages 100--112. {ACM}, 2018.

\bibitem[Hagedorn et~al.(2020)Hagedorn, Lenfers, Koehler, Qin, Gorlatch, and
  Steuwer]{DBLP:journals/pacmpl/HagedornLKQGS20}
Bastian Hagedorn, Johannes Lenfers, Thomas Koehler, Xueying Qin, Sergei
  Gorlatch, and Michel Steuwer.
\newblock Achieving high-performance the functional way: a functional pearl on
  expressing high-performance optimizations as rewrite strategies.
\newblock \emph{Proc. {ACM} Program. Lang.}, 4\penalty0 ({ICFP}):\penalty0
  92:1--92:29, 2020.

\bibitem[Hennessy and Patterson(2019)]{DBLP:journals/cacm/HennessyP19}
John~L. Hennessy and David~A. Patterson.
\newblock A new golden age for computer architecture.
\newblock \emph{Commun. {ACM}}, 62\penalty0 (2):\penalty0 48--60, 2019.

\bibitem[Henriksen et~al.(2017)Henriksen, Serup, Elsman, Henglein, and
  Oancea]{DBLP:conf/pldi/HenriksenSEHO17}
Troels Henriksen, Niels G.~W. Serup, Martin Elsman, Fritz Henglein, and
  Cosmin~E. Oancea.
\newblock Futhark: purely functional gpu-programming with nested parallelism
  and in-place array updates.
\newblock In \emph{{PLDI}}, pages 556--571. {ACM}, 2017.

\bibitem[Kelsey(1995)]{DBLP:conf/irep/Kelsey95}
Richard Kelsey.
\newblock A correspondence between continuation passing style and static single
  assignment form.
\newblock In \emph{Intermediate Representations Workshop}, pages 13--23. {ACM},
  1995.

\bibitem[Koehler and Steuwer(2021)]{DBLP:conf/cgo/KoehlerS21}
Thomas Koehler and Michel Steuwer.
\newblock Towards a domain-extensible compiler: Optimizing an image processing
  pipeline on mobile cpus.
\newblock In \emph{{CGO}}, pages 27--38. {IEEE}, 2021.

\bibitem[Lattner and Adve(2004)]{DBLP:conf/cgo/LattnerA04}
Chris Lattner and Vikram~S. Adve.
\newblock {LLVM:} {A} compilation framework for lifelong program analysis {\&}
  transformation.
\newblock In \emph{{CGO}}, pages 75--88. {IEEE} Computer Society, 2004.

\bibitem[Lattner et~al.(2021)Lattner, Amini, Bondhugula, Cohen, Davis, Pienaar,
  Riddle, Shpeisman, Vasilache, and Zinenko]{DBLP:conf/cgo/LattnerAB0DPRSV21}
Chris Lattner, Mehdi Amini, Uday Bondhugula, Albert Cohen, Andy Davis,
  Jacques~A. Pienaar, River Riddle, Tatiana Shpeisman, Nicolas Vasilache, and
  Oleksandr Zinenko.
\newblock {MLIR:} scaling compiler infrastructure for domain specific
  computation.
\newblock In \emph{{CGO}}, pages 2--14. {IEEE}, 2021.

\bibitem[Lei{\ss}a et~al.(2015)Lei{\ss}a, K{\"{o}}ster, and
  Hack]{DBLP:conf/cgo/LeissaKH15}
Roland Lei{\ss}a, Marcel K{\"{o}}ster, and Sebastian Hack.
\newblock A graph-based higher-order intermediate representation.
\newblock In \emph{{CGO}}, pages 202--212. {IEEE} Computer Society, 2015.

\bibitem[Lei{\ss}a et~al.(2018)Lei{\ss}a, Boesche, Hack, P{\'{e}}rard{-}Gayot,
  Membarth, Slusallek, M{\"{u}}ller, and
  Schmidt]{DBLP:journals/pacmpl/LeissaBHPMSMS18}
Roland Lei{\ss}a, Klaas Boesche, Sebastian Hack, Ars{\`{e}}ne
  P{\'{e}}rard{-}Gayot, Richard Membarth, Philipp Slusallek, Andr{\'{e}}
  M{\"{u}}ller, and Bertil Schmidt.
\newblock Anydsl: a partial evaluation framework for programming
  high-performance libraries.
\newblock \emph{Proc. {ACM} Program. Lang.}, 2\penalty0 ({OOPSLA}):\penalty0
  119:1--119:30, 2018.

\bibitem[L{\"{u}}cke et~al.(2021)L{\"{u}}cke, Steuwer, and
  Smith]{DBLP:conf/cc/LuckeSS21}
Martin L{\"{u}}cke, Michel Steuwer, and Aaron Smith.
\newblock Integrating a functional pattern-based {IR} into {MLIR}.
\newblock In \emph{{CC}}, pages 12--22. {ACM}, 2021.

\bibitem[Paszke et~al.(2019)Paszke, Gross, Massa, Lerer, Bradbury, Chanan,
  Killeen, Lin, Gimelshein, Antiga, Desmaison, K{\"{o}}pf, Yang, DeVito,
  Raison, Tejani, Chilamkurthy, Steiner, Fang, Bai, and
  Chintala]{DBLP:journals/corr/abs-1912-01703}
Adam Paszke, Sam Gross, Francisco Massa, Adam Lerer, James Bradbury, Gregory
  Chanan, Trevor Killeen, Zeming Lin, Natalia Gimelshein, Luca Antiga, Alban
  Desmaison, Andreas K{\"{o}}pf, Edward Yang, Zach DeVito, Martin Raison,
  Alykhan Tejani, Sasank Chilamkurthy, Benoit Steiner, Lu~Fang, Junjie Bai, and
  Soumith Chintala.
\newblock Pytorch: An imperative style, high-performance deep learning library.
\newblock \emph{CoRR}, abs/1912.01703, 2019.

\bibitem[Paszke et~al.(2021)Paszke, Johnson, Duvenaud, Vytiniotis, Radul,
  Johnson, Ragan{-}Kelley, and Maclaurin]{DBLP:journals/pacmpl/PaszkeJDVRJRM21}
Adam Paszke, Daniel Johnson, David Duvenaud, Dimitrios Vytiniotis, Alexey
  Radul, Matthew Johnson, Jonathan Ragan{-}Kelley, and Dougal Maclaurin.
\newblock Getting to the point. index sets and parallelism-preserving autodiff
  for pointful array programming.
\newblock \emph{Proc. {ACM} Program. Lang.}, 5\penalty0 ({ICFP}), 2021.

\bibitem[Ragan{-}Kelley et~al.(2013)Ragan{-}Kelley, Barnes, Adams, Paris,
  Durand, and Amarasinghe]{DBLP:conf/pldi/Ragan-KelleyBAPDA13}
Jonathan Ragan{-}Kelley, Connelly Barnes, Andrew Adams, Sylvain Paris,
  Fr{\'{e}}do Durand, and Saman~P. Amarasinghe.
\newblock Halide: a language and compiler for optimizing parallelism, locality,
  and recomputation in image processing pipelines.
\newblock In \emph{{PLDI}}, pages 519--530. {ACM}, 2013.

\bibitem[Rastello and Tichadou(2018)]{SSA-Book}
Fabrice Rastello and Florent~Bouchez Tichadou, editors.
\newblock \emph{SSA-based Compiler Design}.
\newblock Springer, 2018.

\bibitem[Remmelg et~al.(2016)Remmelg, Lutz, Steuwer, and
  Dubach]{DBLP:conf/ppopp/RemmelgLSD16}
Toomas Remmelg, Thibaut Lutz, Michel Steuwer, and Christophe Dubach.
\newblock Performance portable {GPU} code generation for matrix multiplication.
\newblock In \emph{GPGPU@PPoPP}, pages 22--31. {ACM}, 2016.

\bibitem[Reynolds(1997)]{Reynolds1997}
John~C. Reynolds.
\newblock \emph{The Essence of Algol}, pages 67--88.
\newblock Birkh{\"a}user Boston, 1997.

\bibitem[Rompf et~al.(2013)Rompf, Sujeeth, Amin, Brown, Jovanovic, Lee,
  Jonnalagedda, Olukotun, and Odersky]{DBLP:conf/popl/RompfSABJLJOO13}
Tiark Rompf, Arvind~K. Sujeeth, Nada Amin, Kevin~J. Brown, Vojin Jovanovic,
  HyoukJoong Lee, Manohar Jonnalagedda, Kunle Olukotun, and Martin Odersky.
\newblock Optimizing data structures in high-level programs: new directions for
  extensible compilers based on staging.
\newblock In \emph{{POPL}}, pages 497--510. {ACM}, 2013.

\bibitem[Rosen et~al.(1988)Rosen, Wegman, and Zadeck]{DBLP:conf/popl/RosenWZ88}
Barry~K. Rosen, Mark~N. Wegman, and F.~Kenneth Zadeck.
\newblock Global value numbers and redundant computations.
\newblock In \emph{{POPL}}, pages 12--27. {ACM} Press, 1988.

\bibitem[Scholz(1996)]{DBLP:conf/ifl/Scholz96}
Sven{-}Bodo Scholz.
\newblock On programming scientific applications in {SAC} - {A} functional
  language extended by a subsystem for high-level array operations.
\newblock In \emph{Implementation of Functional Languages}, volume 1268 of
  \emph{Lecture Notes in Computer Science}, pages 85--104. Springer, 1996.

\bibitem[Steele(1976)]{lambda-declarative}
Guy~Lewis Steele.
\newblock Lambda: The ultimate declarative.
\newblock volume 379, 1976.

\bibitem[Steele and Sussman(1976)]{lambda-imperative}
Guy~Lewis Steele and Gerald~Jay Sussman.
\newblock Lambda: The ultimate imperative.
\newblock volume 353, 1976.

\bibitem[Steuwer et~al.(2015)Steuwer, Fensch, Lindley, and
  Dubach]{DBLP:conf/icfp/SteuwerFLD15}
Michel Steuwer, Christian Fensch, Sam Lindley, and Christophe Dubach.
\newblock Generating performance portable code using rewrite rules: from
  high-level functional expressions to high-performance opencl code.
\newblock In \emph{{ICFP}}, pages 205--217. {ACM}, 2015.

\bibitem[Steuwer et~al.(2017)Steuwer, Remmelg, and
  Dubach]{DBLP:conf/cgo/SteuwerRD17}
Michel Steuwer, Toomas Remmelg, and Christophe Dubach.
\newblock Lift: a functional data-parallel {IR} for high-performance {GPU} code
  generation.
\newblock In \emph{{CGO}}, pages 74--85. {ACM}, 2017.

\end{thebibliography}

\end{document}